\documentclass[useAMS,usenatbib]{mnras}
\usepackage{aas_macros}
\usepackage{graphicx}
\usepackage{amsmath,amssymb}
\usepackage[T1]{fontenc}
\usepackage{aecompl}
\usepackage{times}
\title[Baryons in the Cosmic Web -- II]{Baryons in the Cosmic Web of IllustrisTNG -- II: the Connection among Galaxies, Halos, their Formation Time and their Location in the Cosmic Web}
\author[D. Martizzi et al.]{\parbox[t]{\textwidth}{Davide Martizzi$^{1,2}$\thanks{E-mail: davide.martizzi@nbi.ku.dk}, 
Mark Vogelsberger$^{3}$, Paul Torrey$^{4}$, Annalisa Pillepich$^{5}$, Steen H. Hansen$^{1}$, Federico Marinacci$^{6}$, Lars Hernquist$^{7}$} \\ \\
$^{1}$DARK, Niels Bohr Institute, University of Copenhagen, 2100 Copenhagen, Denmark \\
$^{2}$Department of Astronomy and Astrophysics, University of California, Santa Cruz, CA 95064, USA \\
$^{3}$MIT Kavli Institute for Astrophysics \& Space Research, Cambridge, MA, 02139, USA \\
$^{4}$Department of Astronomy, University of Florida, 211 Bryant Space Sciences Center, Gainesville, FL 32611 USA \\
$^{5}$Max-Planck-Institut f\"{u}r Astronomie, K\"{o}nigstuhl 17, 69117 Heidelberg, Germany \\
$^{6}$Department of Physics \& Astronomy, University of Bologna, via Gobetti 93/2, 40129 Bologna, Italy \\
$^{7}$Harvard-Smithsonian Center for Astrophysics, 60 Garden Street, Cambridge, MA 02138, USA \\
}

\begin{document}

\maketitle

\begin{abstract}
The connections among galaxies, the dark matter halos where they form and the properties of the large-scale Cosmic Web still need to be completely disentangled. We use the cosmological hydrodynamical simulation TNG100 of the IllustrisTNG suite to quantify the effects played by the large-scale density field and the Cosmic Web morphology on the relation between halo mass and galaxy stellar mass. We select objects with total dynamical mass in the range $\geq 6.3\times 10^{10} h ^{-1}\, M_{\odot}$ up to a few $10^{14} h^{-1} \, M_{\odot}$ between redshift $z=4$ and redshift $z=0$. A Cosmic Web class (knot, filament, sheet, void) is assigned to each region of the volume using a density field deformation tensor-based method. We find that galaxy stellar mass strongly correlates with total dynamical mass and formation time, and more weakly with large-scale overdensity and Cosmic Web class. The latter two quantities correlate with each other, but are not entirely degenerate. Furthermore, we find that at fixed halo mass, galaxies with stellar mass lower than the median value are more likely to be found in voids and sheets, whereas galaxies with stellar mass higher than the median are more likely to be found in filaments and knots. Finally, we find that the dependence on environment is stronger for satellites than for centrals, and discuss the physical implications of these results.
\end{abstract}


\section{Introduction} \label{sec:intro}

The current theoretical scenario for cosmic structure formation postulates that galaxies form from baryon reservoirs within dark matter halos \citep{1978MNRAS.183..341W}. On the other hand, the formation of dark matter halos is determined by the evolution of small perturbations in the primordial density field growing due to self-gravity and eventually decoupling from the Hubble flow \citep{1980PhRvD..22.1882B,1984PThPS..78....1K}. The connection between dark matter halo and observed galaxies has been the subject of numerous studies during the last 50 years, a field that has been recently summarised in an excellent review by \cite{2018ARA&A..56..435W}. 

The fully non-linear evolution of cosmic dark matter density fields, their baryonic counterpart, and galaxy formation in statistically representative volumes of the Universe can now be studied with cosmological hydrodynamical numerical simulations \citep{2014MNRAS.444.1518V, 2015MNRAS.446..521S, 2016MNRAS.463.3948D,2018MNRAS.475..648P}, or semi-analytical and empirical models built upon dark matter cosmological simulations \citep{doi:10.1146/annurev-astro-082812-140951,2018arXiv180607893B}. These methods can be used to model {the mean relation between the stellar mass of galaxies and the mass of their host halos, the abundance of galaxies and 2-point galaxy clustering. }

Despite the multiple successes of these approaches to study galaxy formation, several details of the connection of galaxies, halos and the large-scale density field still have to be understood. In fact, many galaxy properties, such as stellar masses, stellar ages, metallicities, quantitative measures of morphology, star-formation rates, etc., exhibit a large scatter at fixed mass scale whose origin and connection to the dark matter halo and the large-scale structure still needs to be disentangled. For instance, simulations show that, at fixed halo mass, the clustering of dark matter halos depends on secondary properties (such as formation time, concentration and spin), a phenomenon called halo assembly bias  \citep{2001PhDT.........7W, 2005MNRAS.363L..66G, 2007MNRAS.377L...5G, 2007MNRAS.374.1303C}. The concept of assembly bias has been also extended to galaxies and is still being studied \citep{2014MNRAS.443.3044Z}.

From an observational viewpoint, it was established that galaxy properties correlate with their local environment  \citep{1998ApJ...499..589H,2003ApJ...585L...5H,2004MNRAS.348.1355B,2005ApJ...634..833C,2005ApJ...629..143B,2006PASP..118..517B,2007ApJ...664..791B,2008MNRAS.383.1058C}. More recent work based on broad galaxy surveys has allowed multiple groups to begin investigating the correlation of galaxy properties with their large-scale environment \citep{2013ApJS..206....3S, 2013ApJ...776...71C, 2015MNRAS.448.3665E, 2015MNRAS.448.1483L, 2017MNRAS.466..228E, 2017MNRAS.467L...6P, 2017ChA&A..41..302W, 2018MNRAS.477.3136S, 2018ApJ...854...30P}. The main lesson learned from these studies is that the role of environment needs to also be taken into account in theoretical models that aim to connect galaxy to halo properties. 

The last decade witnessed a huge development of the branch of numerical cosmology that studies the connection between the large-scale Cosmic Web produced by the growth of primordial perturbations, dark matter halos that form within this Web, and galaxies. A series of studies focused on identifying the properties of halos that are related to the morphology of the local Cosmic Web \citep{2011MNRAS.418.2493P,2012MNRAS.421L.137L,2014MNRAS.443.1090F,2015SSRv..193....1J,2018MNRAS.476.4877M,2019MNRAS.483.2101G}. This work was quickly followed by papers focusing on the general connection between galaxies and the Cosmic Web \citep{2015MNRAS.446.1458M,2016arXiv160707881A,2016MNRAS.462..448G,2018MNRAS.474..547K,2018MNRAS.479..973C}, with particular emphasis placed on explaining the origin of the alignment of galaxy/halo spins with cosmic filaments \citep{2012MNRAS.427.3320C,2014MNRAS.444.1453D,2014MNRAS.445L..46W,2015MNRAS.454.2736C,2015MNRAS.448.3391C,2017MNRAS.472.1163C,2019MNRAS.483.3227K,2019arXiv190209797K}. However, this field is young and investigation of the effects of large-scale environment on galaxy populations is still incomplete. 

Cosmological hydrodynamical simulations for galaxy physics are one of the most powerful tools to study the combination of assembly bias and large-scale environmental effects on the formation and evolution of galaxies, because they naturally reproduce the emergence of both halos and galaxies from the large-scale scale density field. Therefore, assembly bias and environmental effects naturally emerge from cosmological hydrodynamical simulations. In the best possible scenario, the results of simulations can be used to make predictions for unknown effects to be measured in the real Universe. 

{In recent years, the origin of the scatter in the relationship between the stellar mass of galaxies and that of their host dark matter halos was investigated by multiple groups using numerical simulations. \cite{2017MNRAS.465.2381M} showed that this scatter cannot be entirely explained by invoking halo formation time or concentration as secondary assembly bias parameters. \cite{2019ApJ...871L..21F} explored the role of halo growth rate as a measure of assembly history, and showed that at fixed halo mass ($<10^{12} \, M_{\odot}$) quickly growing halos have lower stellar masses than slowly growing halos.  \cite{2018ApJ...853...84Z} and \cite{2019MNRAS.tmp.2192B} showed that also environment has an effect on the scatter of the relationship between halo mass and galaxy stellar stellar mass. In this paper, we extend previous work and quantify the role played by both assembly history and Cosmic Web environment on the halo mass versus galaxy stellar mass relation. }

Our results are based on the IllustrisTNG cosmological hydrodynamical simulations which were recently publicly released \citep{2019ComAC...6....2N}. Our results rely on the Cosmic Web classification based on the density field deformation tensor that we recently performed \citep[][Paper I hereafter]{2019MNRAS.486.3766M}. Our methods are described in Section~\ref{sec:methods}. The results of the analysis are reported in Section~\ref{sec:results}. Finally, Section~\ref{sec:conclusions} provides a summary and discussion of our findings. 

\section{Methods} \label{sec:methods}

\subsection{IllustrisTNG Simulations}

We analyse the galaxy population in the TNG100 simulation which is part of the IllustrisTNG suite \citep{2018MNRAS.475..676S, 2018MNRAS.480.5113M, 2018MNRAS.477.1206N, 2018MNRAS.473.4077P,2018MNRAS.475..648P, 2018MNRAS.475..624N}. The IllustrisTNG simulations have been performed with an updated version of the methods used for the Illustris simulations \citep{2013MNRAS.436.3031V, 2014MNRAS.438.1985T, 2014MNRAS.444.1518V, 2014Natur.509..177V, 2014MNRAS.445..175G, 2015MNRAS.452..575S}. The IllustrisTNG model for galaxy formation~\citep[][]{2017MNRAS.465.3291W, 2018MNRAS.473.4077P} includes prescriptions for star formation, stellar evolution, chemical enrichment, primordial and metal-line cooling of the gas, stellar feedback with galactic outflows, black hole formation, growth and multimode feedback. Data from the IllustrisTNG simulations is currently publicly available  \citep{2019ComAC...6....2N}.

A series of papers have shown that the IllustrisTNG simulations are generally successful at reproducing observed demographics and structural properties of galaxies, at least within the statistical and systematic uncertainties of the observational results: e.g. the galaxy mass function at multiple redshifts \citep{2018MNRAS.473.4077P}, the stellar mass content of massive halos \citep{2018MNRAS.475..648P}, galaxy sizes at redshift $0\leq z \leq 2$ \citep{2018MNRAS.474.3976G}, the galaxy color bi-modality \citep{2018MNRAS.475..624N}, the galaxy mass-metallicity relation \citep{2019MNRAS.484.5587T}, low-redshift galaxy star-formation rates \citep{2019MNRAS.485.4817D}, and the detailed properties of unusual galaxies \citep{2018MNRAS.480L..18Z,2019MNRAS.483.1042Y}. In this paper, we re-visit the stellar masses of galaxies in TNG100 and analyse them in connection to the properties of the Cosmic Web. 

\subsection{Cosmic Web Classification}\label{sec:classification}

In Paper I, we analysed the Cosmic Web in the TNG100 simulations and measured the mass fraction of all gas phases from redshift $z=8$ to redshift $z=0$. These results were based on our own Cosmic Web classification tool based on the deformation tensor of the density field (i.e. its Hessian matrix; \citealt{2007ApJ...655L...5A, 2007MNRAS.375..489H, 2009MNRAS.393..457S, 2009MNRAS.396.1815F, 2017ApJ...838...21Z,2018MNRAS.473...68C}). This method assigns a physically-based web class to each region of the simulation at each redshift. First, the total density field is computed via cloud-in-cell interpolation into a regular $512^3$ Cartesian grid. Then, the density field is smoothed with a Gaussian smoothing kernel of radius $R_{\rm G} = 8 \, h^{-1}{\rm Mpc}$. The deformation tensor of the density field is computed at each node of the Cartesian grid and then diagonalised. Based on the eigenvalues of the deformation tensor it is possible to assign a Cosmic Web class to each region of the volume: knots are gravitationally collapsed structures along 3 axes and have 3 eigenvalues larger than $\lambda_{\rm th}=0.3$; filaments are gravitationally collapsed structures along 2 axes and have 2 eigenvalues larger than $\lambda_{\rm th}=0.3$; sheets are gravitationally collapsed structures along 1 axis and have 1 eigenvalue larger than $\lambda_{\rm th}=0.3$; voids are regions that did not undergo gravitational collapse and do not have any eigenvalue larger than $\lambda_{\rm th}=0.3$. The threshold value $\lambda_{\rm th}=0.3$ has been chosen to provide results consistent with previous literature: details and tests are reported in Paper I. 

The Cosmic Web classification method we adopt is ideal to study the role of environment on scales larger than the smoothing scale $R_{\rm G} = 8 \, h^{-1}{\rm Mpc}$, i.e. at cosmological scales. The structure of the Cosmic Web at smaller scales is smoothed out, so that we can focus on the effect of environment at linear and marginally non-linear scales. This is intentional, because we want to separate environmental effects taking place inside and near dark matter halos from those induced by cosmological evolution in the large-scale Cosmic Web. 

\subsection{Galaxy Selection and Merger Trees}

Due to finite numerical resolution, not all the galaxies in the simulation volume can be resolved with a large number of particles/cells. The dark matter particle mass of TNG100 is $m_{\rm dm} = 7.46 \times 10^6 \, {\rm M}_{\odot}$, whereas the initial baryonic mass particle is $ m_{\rm bar}=1.39 \times 10^6 \, {\rm M}_{\odot}$. As the gaseous component was evolved on a moving mesh, the code kept the gas mass resolution within a factor of 2 from this initial value. The spatial resolution is $\sim 1 h^{-1} \, {\rm kpc}$ (see Appendix of \citealt{2019ComAC...6....2N}). 

In this paper, we focus on the relation between total halo mass and stellar mass, and its scatter. {Throughout the paper, instead of using the face value of the total halo mass computed by the {\sc subfind} algorithm, we use $M_{\rm peak}$, the peak (maximum) mass reached by a gravitationally bound object across during past history, computed by including all matter components: dark matter, gas and stars.  This can be interpreted as the dynamical mass of a self-gravitating (sub)halo or galaxy, and it is robust against environment-related events that may cause the (sub)halo mass of satellites to decrease, e.g. stripping processes \citep[e.g.][]{2016MNRAS.460.3100C,2017MNRAS.465.2381M}.} 

For a given halo mass, one typically finds a range of stellar masses. To make sure that our results are not influenced by the finite numerical resolution, we make sure that the halo mass versus stellar mass is thoroughly sampled. At each redshift, we use the {\sc subfind} group catalogues available for TNG100 \citep{2005Natur.435..629S}. We only select objects whose total dynamical mass is $M_{\rm peak}\geq6.3\times 10^{10} \, h^{-1}{\rm M_{\odot}}$. For total masses below this threshold, the median stellar mass expected is $M_{*}<0.001\,M_{\rm peak}\lesssim 60\,m_{\rm bar}$ \citep[e.g.][]{2018arXiv180607893B}, but can be as low as $M_{*}<0.0005\,M_{\rm peak}\lesssim 10\,m_{\rm bar}$, i.e. the regime in which galaxies are not well numerically resolved. In what follows, we consider both central as well as satellite gravitationally-bound haloes/galaxies.

For each object selected at redshift $z_{\rm sel}$, we use the {\sc SubLink} merger trees available for TNG100 \citep{2015MNRAS.449...49R} to identify its formation redshift $z_{\rm form}$, i.e. the redshift at which the total mass was half of the maximum mass of the object along the main branch of the merger tree. This choice allows us to generalise the definition of formation redshift from central to satellite galaxies, which can experience significant mass reduction due to stripping processes. In particular, for satellite galaxies, the halo formation time is by construction larger than the infall time. For each object, we also identify whether they are centrals or satellites embedded in a larger halo, using the information contained in the {\sc subfind} catalogues. 

Finally, at each selected redshift, we use the $512^3$ Cartesian grid with Cosmic Web classes produced by our classification code (Section~\ref{sec:classification}) to assign a class $W$ to each galaxy in the simulation volume, according to its location: (I) galaxies in knots receive $W=3$, (II) galaxies in filaments receive $W=2$, (III) galaxies in sheets receive $W=1$, and (IV) galaxies in voids receive $W=0$. Furthermore, we store the large-scale total matter overdensity value $\delta_8=\rho/\bar{\rho}-1$ (smoothed with a Gaussian kernel of radius $R_{\rm G} = 8 \, h^{-1}{\rm Mpc}$) at the galaxy position and at the selection redshift, by accounting for all matter in the box i.e. dark matter, gas, stars, black holes.

Equipped with formation redshift $z_{\rm form}$, large-scale overdensity $\delta_8$, Cosmic Web class $W$, and central versus satellite indicator for each galaxy selected at redshift $z_{\rm sel}$, we analyse how the properties of the total mass versus stellar mass relation are influenced by these variables. 

\section{Results} \label{sec:results}

\subsection{The Total Mass versus Stellar Mass Relation and the Cosmic Web}\label{sec:res-nq}

\begin{figure*}
\begin{center}
\includegraphics[width=0.345\textwidth]{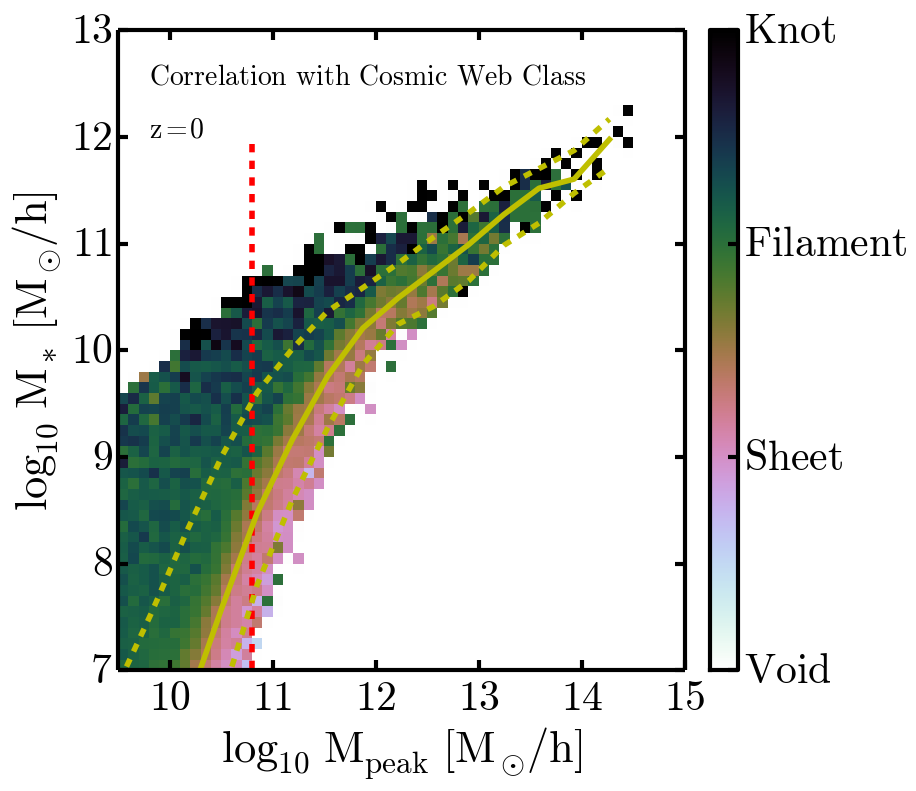}
\includegraphics[width=0.345\textwidth]{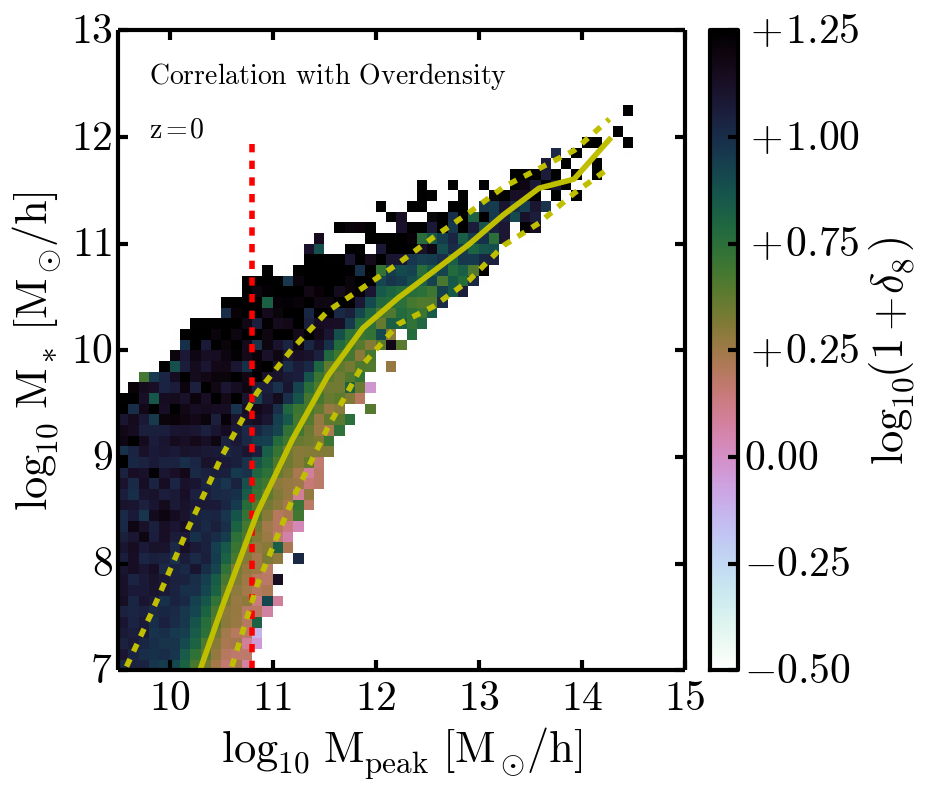}
\includegraphics[width=0.345\textwidth]{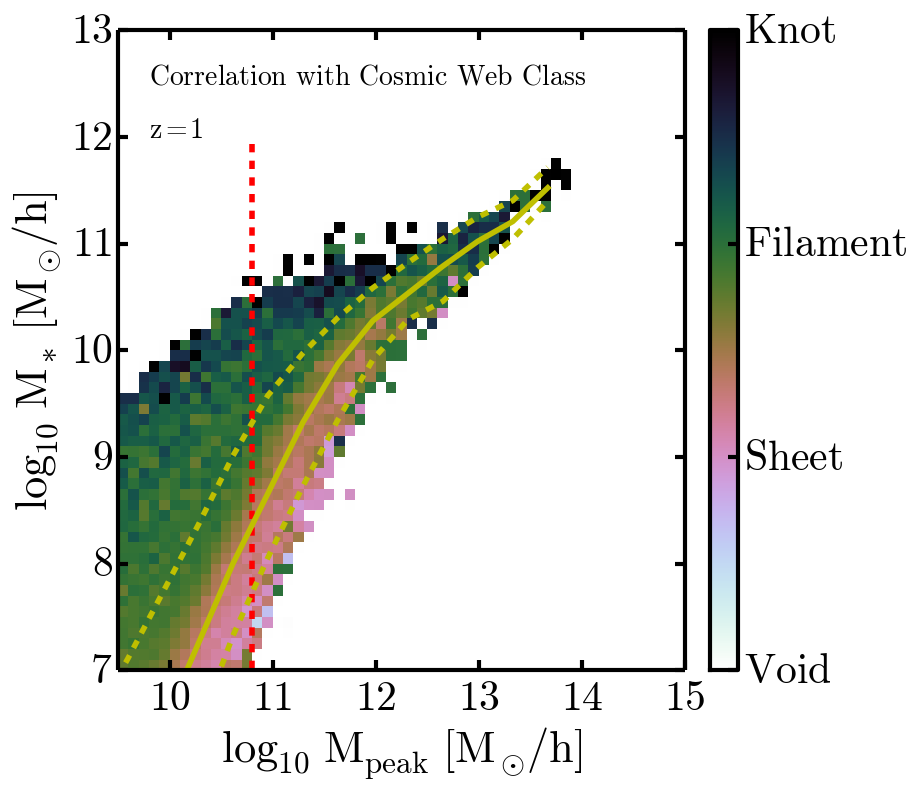}
\includegraphics[width=0.345\textwidth]{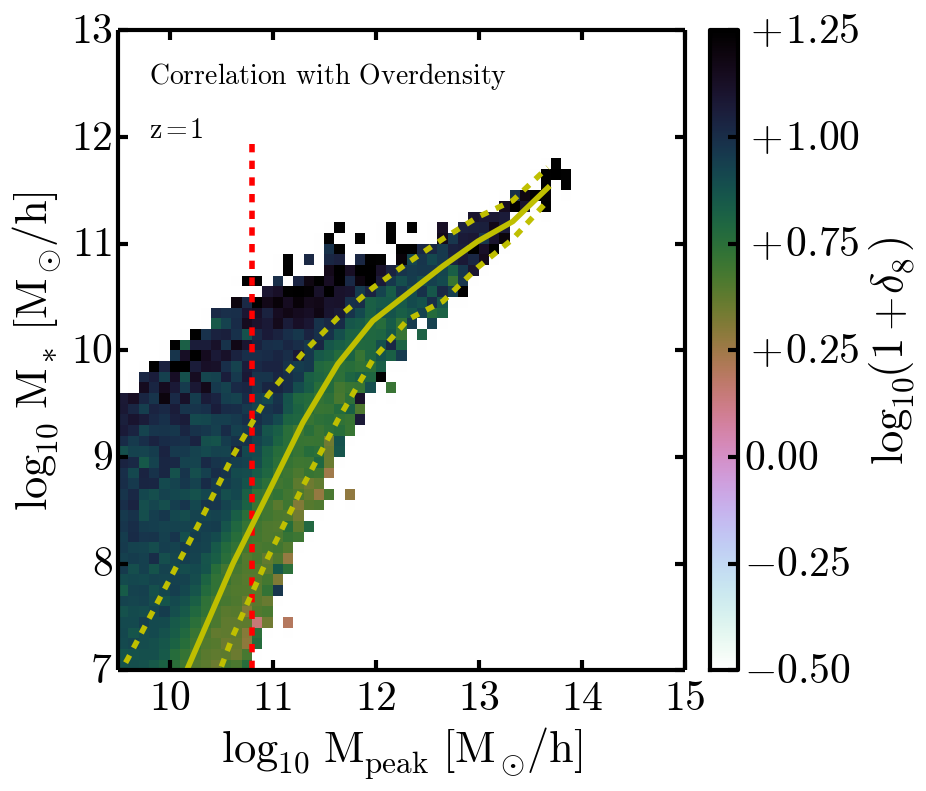}
\includegraphics[width=0.345\textwidth]{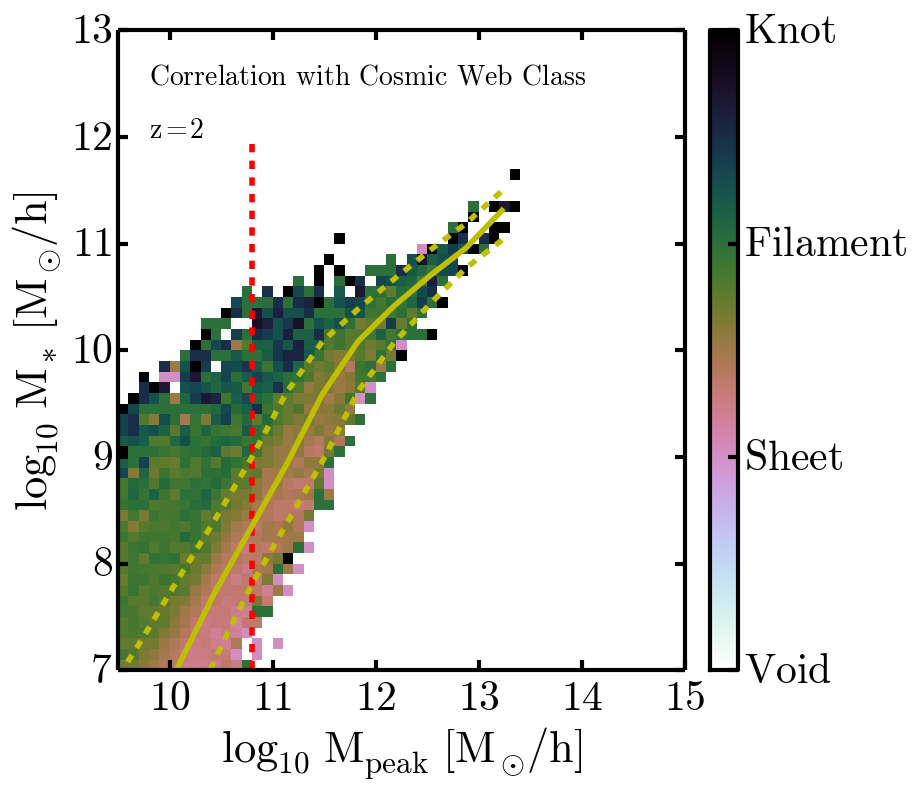}
\includegraphics[width=0.345\textwidth]{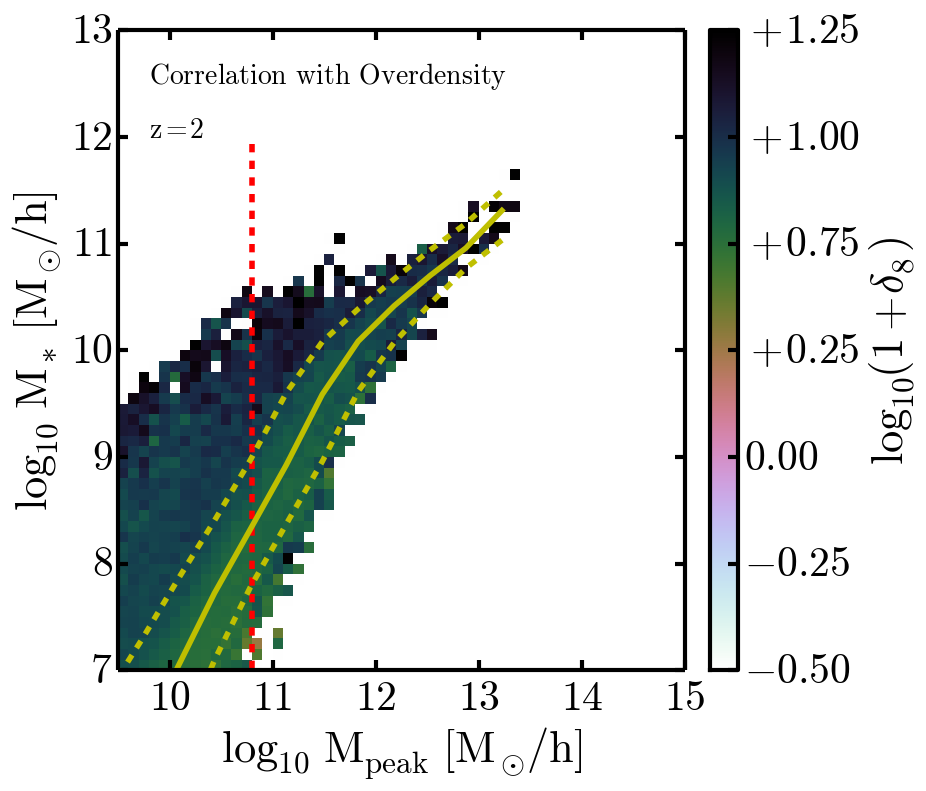}
\includegraphics[width=0.345\textwidth]{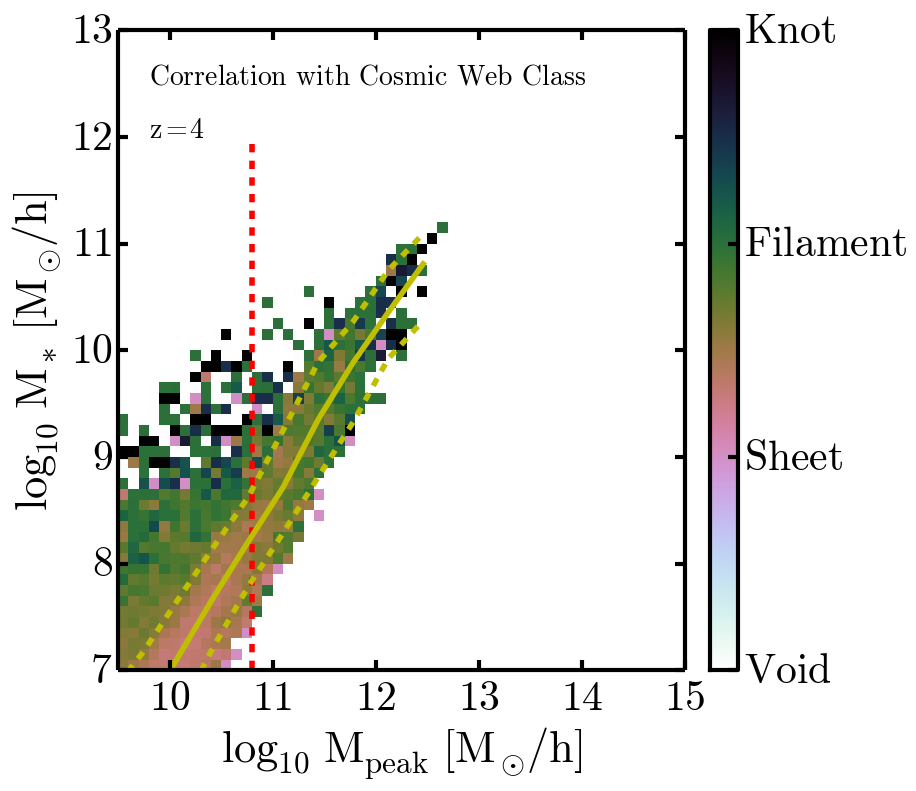}
\includegraphics[width=0.345\textwidth]{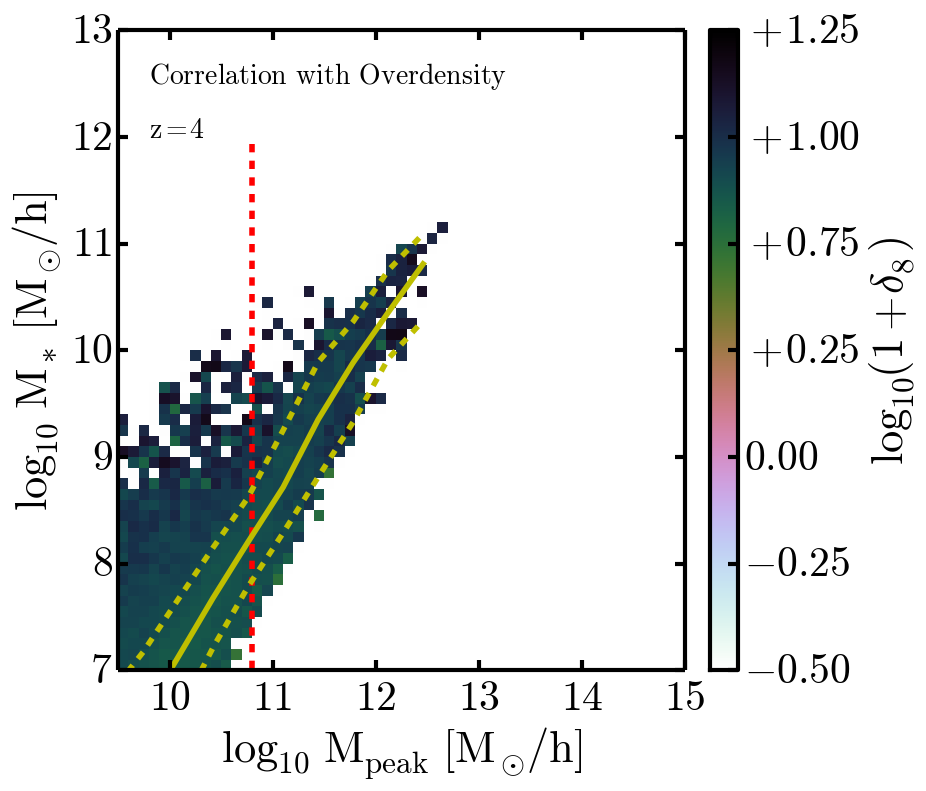}
\end{center}
\caption{TNG100 simulation: total mass versus galaxy stellar mass at redshift $z=0,1,2,4$ going from top to bottom. The total mass of each halo has been calculated using the {\sc subfind} algorithm. The color scale represent the median Cosmic Web class (left panels) and large-scale overdensity (right panels) in each $\log_{10}M_{\rm peak}-\log_{10}M_*$ bin. A color is assigned to each pixel only when at least one galaxy falls into it. The yellow solid line represent the median stellar mass versus total mass relation for the whole population at each redshift, whereas the dashed yellow lines represents running 5 and 95 percentiles, respectively. The red dashed-line represents the cut we apply in the selection of galaxies/halos in our quantitative analysis. The scatter in the total mass versus stellar mass relation depends on the location in the Cosmic Web (left panels) and the large-scale overdensity $\delta_8$ (right panels).}\label{fig:stellarmass_evo_comparison}
\end{figure*}

We begin our discussion of the results by showing the total mass versus stellar mass relation at redshifts $z=0,1,2, \, {\rm and} \, 4$ in a novel way in Figure~\ref{fig:stellarmass_evo_comparison}. The left panels are 2-d histograms in the $\log_{10}M_{\rm peak}-\log_{10}M_{*}$ plane that show the median Cosmic Web class in each bin (\textit{not} the galaxy counts). These plots reveal that in all considered snapshots the position of galaxies in the $\log_{10}M_{\rm peak}-\log_{10}M_{*}$ depends on the Cosmic Web class $W$, i.e. on the galaxy location in the Cosmic Web. At each given total mass, the most massive galaxies tend to be in knots and filaments. Conversely, lower mass galaxies tend to be in sheets and voids. 

The correlation of galaxy properties with the large-scale density field has been recently quantified \citep{2015MNRAS.448.3665E,2015MNRAS.448.1483L,2017MNRAS.466..228E,2017MNRAS.467L...6P,2017ChA&A..41..302W,2018MNRAS.477.3136S,2018ApJ...854...30P}. Such correlation is also found in TNG100, as shown by the right panels of Figure~\ref{fig:stellarmass_evo_comparison}, which are similar to the left panels, but show the median $\log_{10}(1+\delta_8)$ in each $\log_{10}M_{\rm peak}-\log_{10}M_{*}$ bin. These plots show that at fixed total mass, the most massive galaxies tend to be in high density large-scale environments, whereas low-mass galaxies tend to be in lower density environments. 

Figure~\ref{fig:stellarmass_evo_comparison} does not provide quantitative information on whether stellar masses correlate more strongly with Cosmic Web class $W$ or large-scale overdensity $\delta_8$. In principle, $W$ and $\delta_8$ could be partially degenerate, because $W$ contains information about the local shape of the density field. For this reason, before drawing conclusions, we need to quantitatively assess whether the two variables really have any predictive power on the position of galaxies in the $\log_{10}M_{\rm peak}-\log_{10}M_{*}$ plane. 

\begin{table*}
\centering
\caption{ The normalised covariance matrix between total mass $\log_{10}M_{\rm peak}$, halo formation time $\log_{10}(1+z_{\rm form})$, large-scale overdensity $\log_{10}(1+\delta_{8})$ and Cosmic Web class $W$ at multiple redshifts. Values close to 1 or -1 indicate a higher correlation between the variables. The large-scale overdensity $\delta_8$ and the Cosmic Web class are not entirely correlated and the correlation strength decreases from low to high redshift.}\label{tab:covariance}
{\bfseries Covariance Matrix of Halo Properties}
\makebox[\linewidth]{
\begin{tabular}{llcccc}
\hline
\hline
\multicolumn{6}{l}{Redshift $z=0.0$} \\ 
\hline
 & & $\log_{10}M_{\rm peak}$ & $\log_{10}(1+z_{\rm form})$ & $\log_{10}(1+\delta_{8})$ & $W$\\
 & $\log_{10}M_{\rm peak}$ & 1.0000 & -0.0987 & 0.2307 & 0.2000 \\ 
 & $\log_{10}(1+z_{\rm form})$ & -0.0987 & 1.0000 & 0.2495 & 0.2168 \\
 & $\log_{10}(1+\delta_{8})$ & 0.2307 & 0.2495 & 1.0000 & 0.8460 \\
 & $W$ & 0.2000 & 0.2168 & 0.8460 & 1.0000 \\
\hline
\hline
\multicolumn{6}{l}{Redshift $z=1.0$} \\ 
\hline
 & & $\log_{10}M_{\rm peak}$ & $\log_{10}(1+z_{\rm form})$ & $\log_{10}(1+\delta_{8})$ & $W$\\
 & $\log_{10}M_{\rm peak}$ & 1.0000 & -0.1426 & 0.1990 & 0.2039 \\
 & $\log_{10}(1+z_{\rm form})$ & -0.1426 & 1.0000 & 0.1081 & 0.1268 \\
 & $\log_{10}(1+\delta_{8})$ & 0.1990 & 0.1081 & 1.0000 & 0.7145 \\
 & $W$ & 0.2039 & 0.1268 & 0.7145 & 1.0000 \\
\hline
\hline
\multicolumn{5}{l}{Redshift $z=2.0$} \\ 
\hline
 & & $\log_{10}M_{\rm peak}$ & $\log_{10}(1+z_{\rm form})$ & $\log_{10}(1+\delta_{8})$ & $W$\\
 & $\log_{10}M_{\rm peak}$ & 1.0000 & -0.1019 & 0.1854 & 0.1987 \\
 & $\log_{10}(1+z_{\rm form})$ & -0.1019 & 1.0000 & 0.0492 &  0.0737 \\
 & $\log_{10}(1+\delta_{8})$ & 0.1854 & 0.0492 & 1.0000 & 0.6606 \\
 & $W$ & 0.1987 & 0.0737 & 0.6606 & 1.0000 \\
\hline
\hline
\multicolumn{5}{l}{Redshift $z=4.0$} \\ 
\hline
 & & $\log_{10}M_{\rm peak}$ & $\log_{10}(1+z_{\rm form})$ & $\log_{10}(1+\delta_{8})$ & $W$\\
 & $\log_{10}M_{\rm peak}$ & 1.0000 & -0.0914 & 0.1205 & 0.1906 \\
 & $\log_{10}(1+z_{\rm form})$ & -0.0914 & 1.0000 & -0.0355 & 0.0042 \\
 & $\log_{10}(1+\delta_{8})$ & 0.1205 & -0.0355 & 1.0000 & 0.4140 \\
 & $W$ & 0.1906 & 0.0042 & 0.4140 & 1.0000 \\
\hline
\hline
\end{tabular}
}
\end{table*}

\begin{table}
\centering
\caption{ Linear regression coefficients for the scaling of the logarithmic offset of galaxy stellar masses $\Delta_*$ as a function of (normalised) total mass $(\log_{10}M_{\rm peak})/15$, halo formation redshift $\log_{10}(1+z_{\rm form})$, large-scale overdensity $\log_{10}(1+\delta_{8})$ and Cosmic Web class $W$ at multiple redshifts. See equations~\ref{eq:Lall}-\ref{eq:LW}.}\label{tab:params}
{\bfseries Linear model coefficients for $\Delta_*$}
\makebox[\linewidth]{
\begin{tabular}{lrrrr}
\hline
\hline
\multicolumn{5}{l}{Model $\Delta_*(L_{\rm all})$} \\ 
\hline
 Parameter & $z=0$ & $z=1$ & $z=2$ & $z=4$ \\
\hline
 $\Delta_{\rm all}$ & -0.2763 & -0.4173 & -0.4693 & -0.6447 \\ 
 $a_{\rm M}$ & 0.2586 & 0.3233 & 0.2451 & 0.1952 \\ 
 $a_{\rm form}$ & 0.2115 & 0.3338 & 0.4620 & 0.6456 \\ 
 $a_{\rm \delta}$ &  0.0223 & 0.0054 & 0.0015 & -0.0006 \\ 
 $a_{\rm W}$ & 0.0039 & 0.0157 & 0.0111 & 0.0012 \\ 
 \hline
 \hline
\multicolumn{5}{l}{Model $\Delta_*(L_{\rm form})$} \\ 
\hline
 Parameter & $z=0$ & $z=1$ & $z=2$ & $z=4$ \\
\hline
 $\Delta_{\rm form}$ & -0.3351 & -0.4539 & -0.4878 & -0.6465 \\ 
 $b_{\rm M}$ & 0.3418 & 0.3776 & 0.2747 & 0.1975 \\ 
 $b_{\rm form}$ & 0.2366 & 0.3472 & 0.4673 & 0.6459 \\
 \hline
 \hline
\multicolumn{5}{l}{Model $\Delta_*(L_{\rm \delta})$} \\ 
\hline
 Parameter & $z=0$ & $z=1$ & $z=2$ & $z=4$ \\
\hline
 $\Delta_{\rm \delta}$ & -0.1157 & -0.1399 & -0.0982 & -0.0617 \\ 
 $c_{\rm M}$ & 0.1351 & 0.1768 & 0.1238 & 0.0871 \\ 
 $c_{\rm \delta}$ & 0.0409 & 0.0186 & 0.0098 & -0.0025 \\
 \hline
 \hline
\multicolumn{5}{l}{Model $\Delta_*(L_{\rm W})$} \\ 
\hline
 Parameter & $z=0$ & $z=1$ & $z=2$ & $z=4$ \\
\hline
 $\Delta_{\rm W}$ & -0.1469 & -0.1439 & -0.0927 & -0.0571 \\ 
 $d_{\rm M}$ & 0.1664 & 0.1683 & 0.1091 & 0.0754 \\ 
 $d_{\rm W}$ & 0.0522 & 0.0388 & 0.0235 & 0.0030 \\ 
\hline
\hline
\end{tabular}
}
\end{table}

In principle, even the tiniest residual correlation of stellar mass with Cosmic Web location would not be trivial, because it would suggest a connection between galaxy properties and the large-scale deformation tensor of the density field, i.e. the large-scale morphology of the Cosmic Web region where the galaxy lies. 

\subsection{Quantitative Analysis of the residuals in Total Mass versus Stellar Mass Relation}\label{seq:quant_res}

To perform a quantitative analysis of our results, we group the selected galaxies/halos in logarithmic bins in total mass of size $\Delta\log_{10}M_{\rm peak}/M_{\odot} = 0.5$. {In each total mass bin, we compute the median $\log_{10}M_{*}$, which we indicate as $\langle \log_{10}M_{*}|\log_{10}M_{\rm peak}\rangle$. The median trends at each redshift are shown as yellow lines in Figure~\ref{fig:stellarmass_evo_comparison}. These lines are produced by interpolating the values in the logarithmic total mass bins.} Then, for each galaxy/halo we compute the logarithmic offset of its stellar mass with respect to the average relation $\langle \log_{10}M_{\rm peak}|\log_{10}M_{*}\rangle$:
\begin{equation}
 \Delta_* = \frac{\log_{10}M_{*}}{\langle \log_{10}M_{*}|\log_{10}M_{\rm peak}\rangle}-1.
\end{equation}
In practice, $\Delta_*$ quantifies how much the stellar mass deviates from the typical stellar mass expected for galaxies in an object of a given total mass. {This definition decouples the median, non-linear relation between total mass and stellar mass from its scatter, which can then be separately analysed.} The scatter in the total mass versus stellar mass decreases with increasing total mass (Figure~\ref{fig:stellarmass_evo_comparison}; see also Figure 11 of \citealt{2018MNRAS.475..648P}), which suggests that $\Delta_*$ might also depend on $\log_{10}M_{\rm peak}$. Furthermore, Figure~\ref{fig:stellarmass_evo_comparison} suggests that $\Delta_*$ may depend on the large-scale overdensity $\delta_8$ and on the location in the Cosmic Web, quantified by the class $W$. A dependence of $\Delta_*$ on $\delta_8$ and $W$ encapsulates all the environmental effects that can influence a galaxy stellar mass throughout its evolution. Finally, the stellar mass of a galaxy may also depend on the processes that determine the formation of its main progenitor at higher redshift. We parameterise this dependence with the formation redshift $z_{\rm form}$. In our preliminary analysis, we have also tested whether $\Delta_*$ depends on the galaxy star formation rate, but we did not find evidence that this is the case. 

The first quantitative inspection of our sample of simulated galaxies is summarised in Table~\ref{tab:covariance}, which shows the covariance matrix between $\log_{10}M_{\rm peak}$, $\log_{10}(1+z_{\rm form})$, $\log_{10}(1+\delta_{8})$ and $W$ at multiple redshifts. Most of these variables exhibit relatively weak correlation with each other, with the exclusion of $\log_{10}(1+\delta_{8})$ and $W$ whose covariance is $\sim 0.8$ ($\sim 0.4$) at redshift $z=0$ ($z=4$). As discussed in Section~\ref{sec:res-nq}, these two variables partially correlate because they carry information about the density field. However,  $\log_{10}(1+\delta_{8})$ and $W$ are not entirely degenerate and the strength of the correlation quickly decreases from low to high redshift. This point is quite important for our purpose, because we want to use $\log_{10}M_{\rm peak}$, $\log_{10}(1+z_{\rm form})$, $\log_{10}(1+\delta_{8})$ and $W$ as explanatory variables for the stellar mass offset $\Delta_*$.

We perform a quantitative analysis of the dependence of $\Delta_*$ on $\log_{10}M_{\rm peak}$, $\log_{10}(1+z_{\rm form})$, $\log_{10}(1+\delta_8)$ and $W$ by using linear regression analysis at each redshift. First, we perform linear regression of $\Delta_*$ as a function of all the explanatory variables:
\begin{align}\label{eq:Lall}
&\Delta_{\rm *}(L_{\rm all}) = \Delta_{\rm all} + L_{\rm all}, \nonumber \\
&L_{\rm all} = a_{\rm M}\frac{\log_{10}M_{\rm peak}}{15} + a_{\rm form}\log_{10}(1+z_{\rm form}) + \nonumber\\ 
 &\,\,\,\,\,\,\,\,\,\,\,\,\,\,\,\,\,\,a_{\delta}\log_{10}(1+\delta_8) + a_{\rm W}W.
\end{align}
\newline
To assess the predictive power of each explanatory variable, we also consider linear models that only include a subset of them. We always include $\log_{10}M_{\rm peak}$, and perform the following regressions:
\begin{align}
&\Delta_{\rm *}(L_{\rm form}) = \Delta_{\rm form} + L_{\rm form}, \nonumber \\
\label{eq:Lform} &L_{\rm form} = b_{\rm M}\frac{\log_{10}M_{\rm peak}}{15} + b_{\rm form}\log_{10}(1+z_{\rm form}),  \\
 \nonumber \\
&\Delta_{\rm *}(L_{\rm \delta}) = \Delta_{\rm \delta} + L_{\rm \delta}, \nonumber \\
\label{eq:Ldelta} &L_{\rm \delta} = c_{\rm M}\frac{\log_{10}M_{\rm peak}}{15} + c_{\rm \delta}\log_{10}(1+\delta_8), \\ 
 \nonumber \\
&\Delta_{\rm *}(L_{\rm W}) = \Delta_{\rm W} + L_{\rm \delta}, \nonumber \\
\label{eq:LW} &L_{\rm W} = d_{\rm M}\frac{\log_{10}M_{\rm peak}}{15} + d_{\rm W}W.  
\end{align}
The linear regressions are performed with the {\sc scikit-learn} Python library, which also returns a score $R^2 = 1-u/v$, where $u = \sum (\Delta_{\rm *,true}-\Delta_{\rm *,model})^2$ is the sum of residuals, $v = \sum (\Delta_{\rm *,true}-\langle\Delta_{\rm *,true}\rangle)^2$ is the sum of squares, and $\langle \Delta_{\rm *,true}\rangle$ is the mean of the true values of $\Delta_*$. $u$ quantifies the accuracy of the linear model, whereas $v$ is proportional to the variance in the data that the model should reproduce. A perfect linear model will have $u=0$ (zero residuals), score $R^2=1$, and will fully reproduce the variance observed in $\Delta_{\rm *,true}$ as a function of the explanatory variables. Any imperfect model will have $u\neq 0$, and yield partially inaccurate prediction which produce artificial variance in the predicted values $\Delta_{\rm *,model}$. This spurious variance can be quantified by the ratio $u/v$. On the other hand, the fraction of the variance in $\Delta_{\rm *,true}$ that can be correctly captured by the model is given by the linear correlation score $R^2=1-u/v$. Low values of $R^2$ imply that the model is inaccurate (very large residual), or that $\Delta_*$ depends weakly on the explanatory variable, leaving a large fraction of the variance unexplained by the model. {All the linear regressions were performed by assigning an equal weight to each galaxy. The consequences of this choice are discussed in Subsection~\ref{sec:mcut}.} 

\begin{figure*}
\begin{center}
\includegraphics[width=0.9\textwidth]{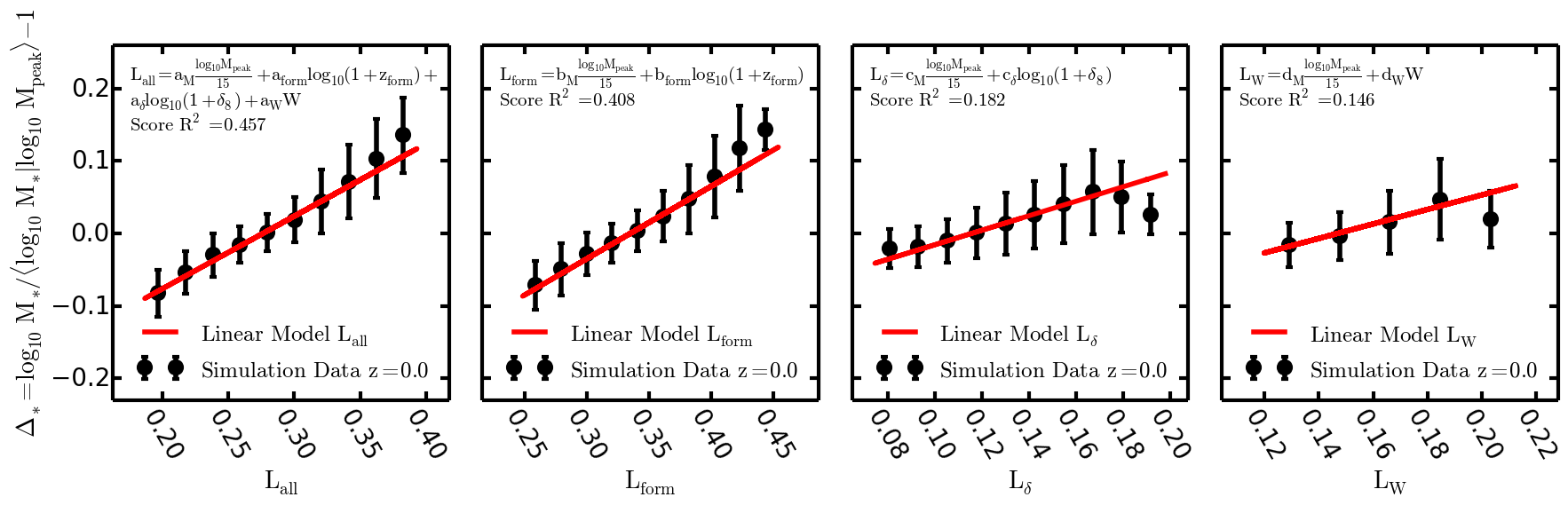}
\includegraphics[width=0.9\textwidth]{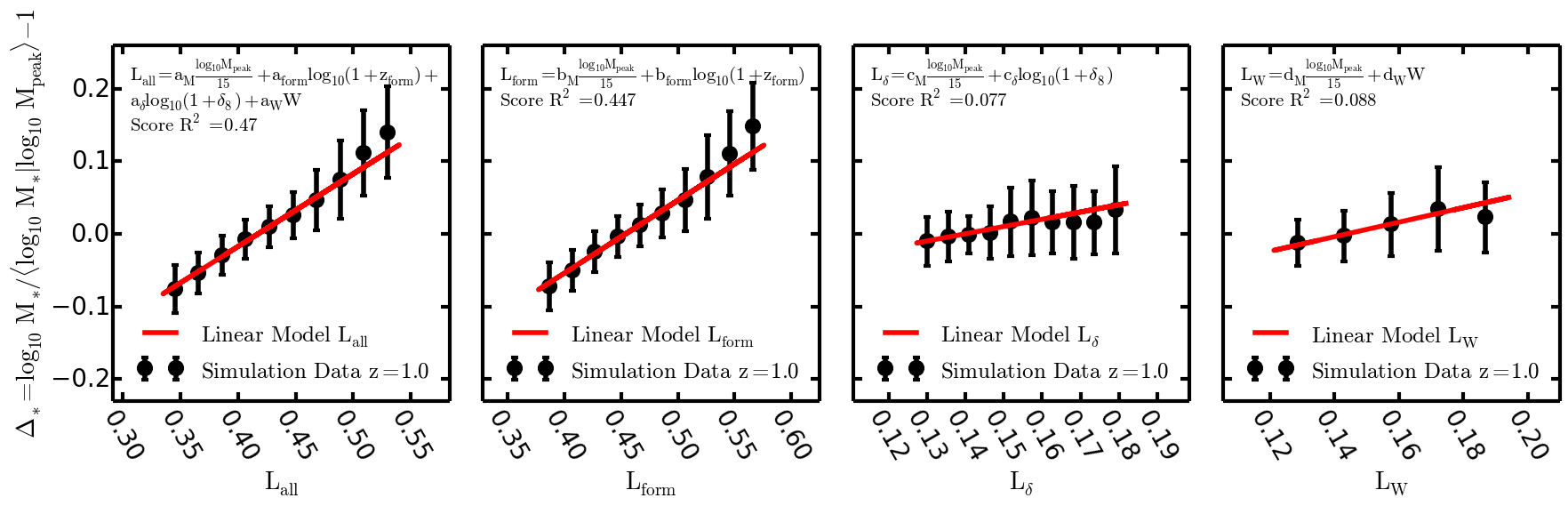}
\includegraphics[width=0.9\textwidth]{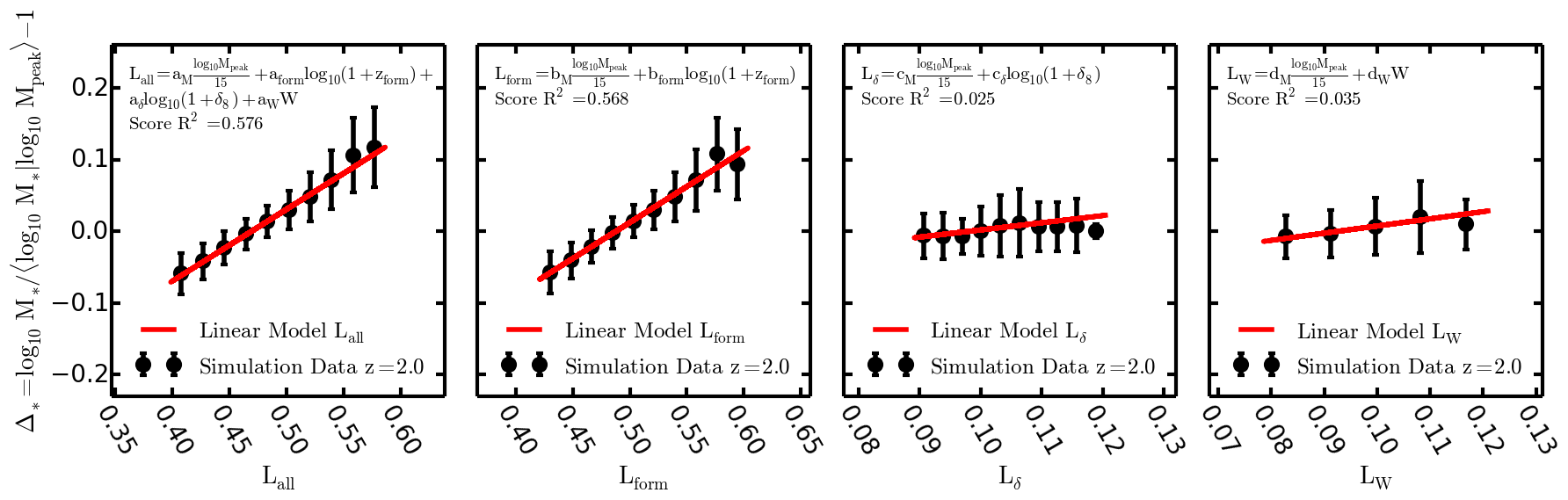}
\includegraphics[width=0.9\textwidth]{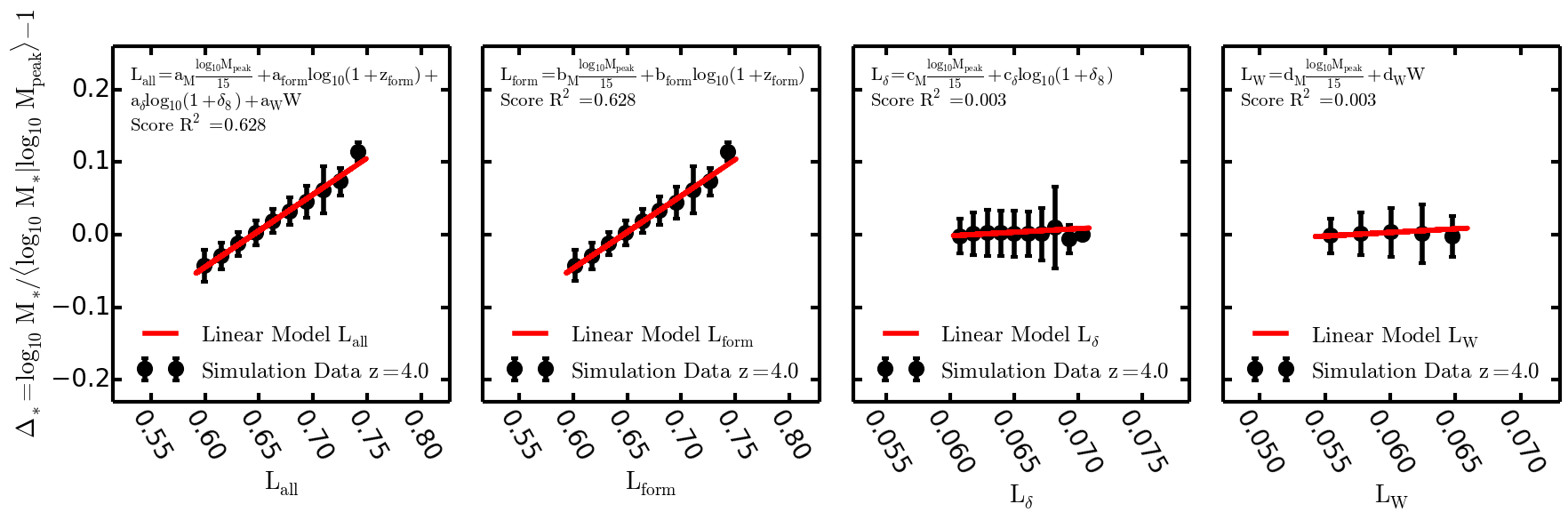}
\end{center}
\caption{Logarithmic offset $\Delta_*$ of $\log_{10}M_*$ with respect to the median measured at fixed total mass $\langle \log_{10}M_{*}|\log_{10}M_{\rm peak}\rangle$. $\Delta_*$ is plotted as as a function of the linear variables $L_{\rm all}, L_{\rm form}, L_{\rm \delta}, L_{\rm W}$ defined in equations~\ref{eq:Lall}, \ref{eq:Lform}, \ref{eq:Ldelta}, \ref{eq:LW}, respectively. The black points with error bars represent data from TNG100 (mean $\Delta_*$ and $1\sigma$ scatter within each bin). The red lines represent results of linear regressions of $\Delta_*$ with respect to the linear explanatory variables reported on the x-axes. Each row represents a different redshift, $z=0, 1, 2, 4$ going from the top row to the bottom row, respectively. The linear regression score $R^2$ is reported in each panel. At all redshifts the combination of total mass $M_{\rm peak}$ and formation redshift $z_{\rm form}$ is the strongest predictor for $\Delta_*$. The role of environmental variables related to the Cosmic Web (large-scale overdensity $\delta_8$ and Cosmic Web class $W$) emerges at $z<2$, and becomes somewhat stronger as redshift decreases. {Nonetheless, the large-scale overdensity $\delta_8$ and the Cosmic Web class $W$ are weaker predictors of $\Delta_*$ than the formation redshift $z_{\rm form}$}.} \label{fig:deltamstar_linear}
\end{figure*}

To assess the predicting power of each explanatory variable we compare the scores for each regression at each redshift. The coefficients of the linear models measured via linear regression are reported in Table~\ref{tab:params}, whereas the values of the scores are reported in Figure~\ref{fig:deltamstar_linear}.

Figure~\ref{fig:deltamstar_linear} shows the results of our linear regressions at $z=0, 1, 2, \, {\rm and} \, 4$. The score of the full linear regression $\Delta_*$ versus $L_{\rm all}$ in equation~\ref{eq:Lall} is usually the best, and a linear model appears to provide  excellent fits to the simulations data at all examined redshifts. The linear regression $\Delta_*$ versus $L_{\rm form}$ provides an excellent, but slightly worse fit to the simulation data (lower $R^2$ score) at all redshifts. This implies that $\Delta_*$ most strongly depends on the total mass and on the formation redshift. This might indicate that the final mass of a galaxy is strongly determined by the conditions of the environment where it formed, and to internal processes that are directly or indirectly related to the total mass (e.g. shock heating of the gas, stellar feedback or AGN feedback).

\begin{figure}
\begin{center}
\includegraphics[width=0.45\textwidth]{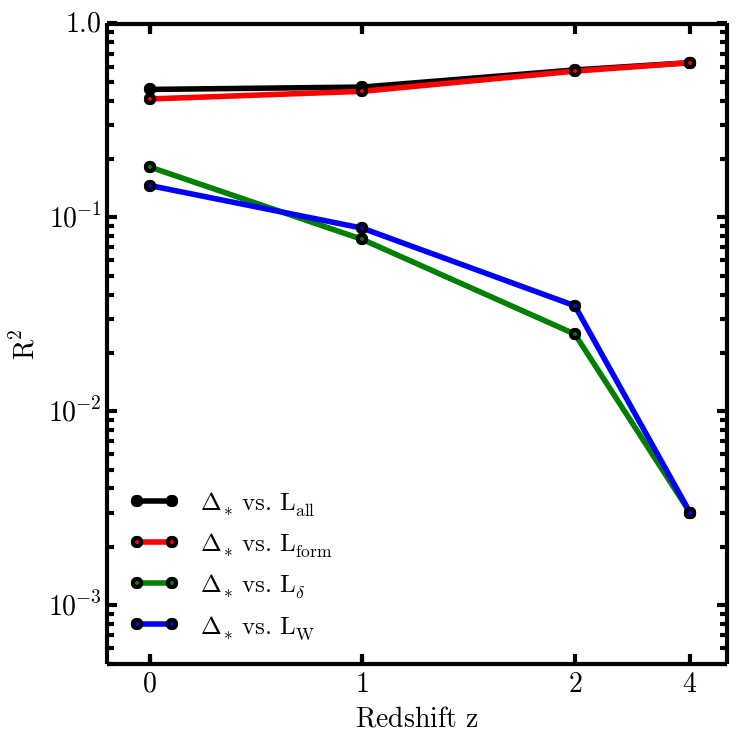}
\end{center}
\caption{Linear correlation scores $R^2$ for the fits to the TNG100 data shown in Figure~\ref{fig:deltamstar_linear}. This scores quantifies the fraction of the variance in $\Delta_*$ that can be explained by each given linear model achieved by linear regression. Our linear models for $\Delta_*$ are excellent fits to the TNG100 simulation data, but become increasingly less predictive from high to low redshift (black line). The variance in $\Delta_*$ can be described well by only including total mass and formation time, but this incomplete model accounts for less than 40\% of the variance at redshift $z\leq 1$ (red line). {Large-scale overdensity and Cosmic Web class account for $\lesssim 5\%$ of the variance in $\Delta_*$ at redshift $z=2$, a fraction that increaseses to $\lesssim 20\%$ at redshift $z=0$ (green and blue lines).} } \label{fig:scores}
\end{figure} 

{Figure~\ref{fig:deltamstar_linear} also shows that total mass and formation redshift are sufficient to explain $\sim 55-60$\% of the measured variance in $\Delta_*$ measured among TNG100 galaxies at redshift $z>2$.} However, this is not the end of the story: at redshift $z<2$ we clearly see an increase in the relevance of the large-scale overdensity $\delta_8$ and of the Cosmic Web class $W$, quantified by the higher $R^2$ scores for the $\Delta_*$ versus $L_{\rm \delta}$ and $\Delta_*$ versus $L_{\rm W}$ regressions, respectively. This evolution can perhaps be better appreciated by looking at Figure~\ref{fig:scores}, which summarises the scores $R^2$ for all the linear regressions we performed in this paper. The combination of large-scale overdensity and Cosmic Web class is able to explain $\lesssim 20\%$ of the total variance in $\Delta_*$ at redshift $z=0$, a fraction that decreases to $\lesssim 5\%$ at $z=2$. {Figure~\ref{fig:scores} also shows that large-scale overdensity and Cosmic Web class have a very similar impact on $\Delta_*$, with differences that do not appear to be significant.}

{Our results seem to also suggest that a significant fraction ($\sim 45-60$\%) of the scatter in the total mass vs. stellar mass relation is unexplained by the linear models considered in this paper at redshift $z\leq 2$. This fact could be the effect of non-linearity, as well as dependence on other stochastic effects that were not considered in our analysis. Non-linearity is also observed in the trend of $\Delta_*$ at the highest values of $L_{\rm \delta}$ and $L_{W}$, which correspond to the most massive halos and the highest large-scale overdensities. Further investigation beyond the scope of this paper will be required to identify the sources of these effects.}

{To summarise, we conclude that information on the large-scale density field can be important when trying to explain the scatter in the total mass versus stellar mass relation, but its magnitude correlates more strongly with halo formation redshift.} 

\subsection{Central versus Satellite Galaxies}

We have so far discussed the analysis of the whole population of galaxies. However, galaxies that survive at the centre of dark matter halos typically have very different histories compared to their satellites. On average, centrals selected at low redshift formed earlier, are more massive and have undergone multiple major and minor mergers. On the other end, on average, satellites form and are accreted onto a larger dark matter halo later in cosmic history, and they undergo multiple environmental effects, such as stripping of their dark matter, gaseous and stellar content. For this reason, one should expect centrals and satellites to have different dependencies on the properties of their halo, and on environmental variables such as local large-scale overdensity of Cosmic Web class.

Driven by these arguments, we separate the sample of galaxies selected at redshift $z=0$ in centrals versus satellites, then we repeat the linear regression analysis we discussed above. The TNG100 halo catalogues provide a list of friends-of-friends (FOF) groups and a separate list of {\sc subfind} objects, which are contained within the FOF groups. Here, centrals are defined as galaxies that sit at the centre of a host FOF group. Satellites are galaxies that belong to the same FOF group, but are not central. 

\begin{figure*}
\begin{center}
\includegraphics[width=0.87\textwidth]{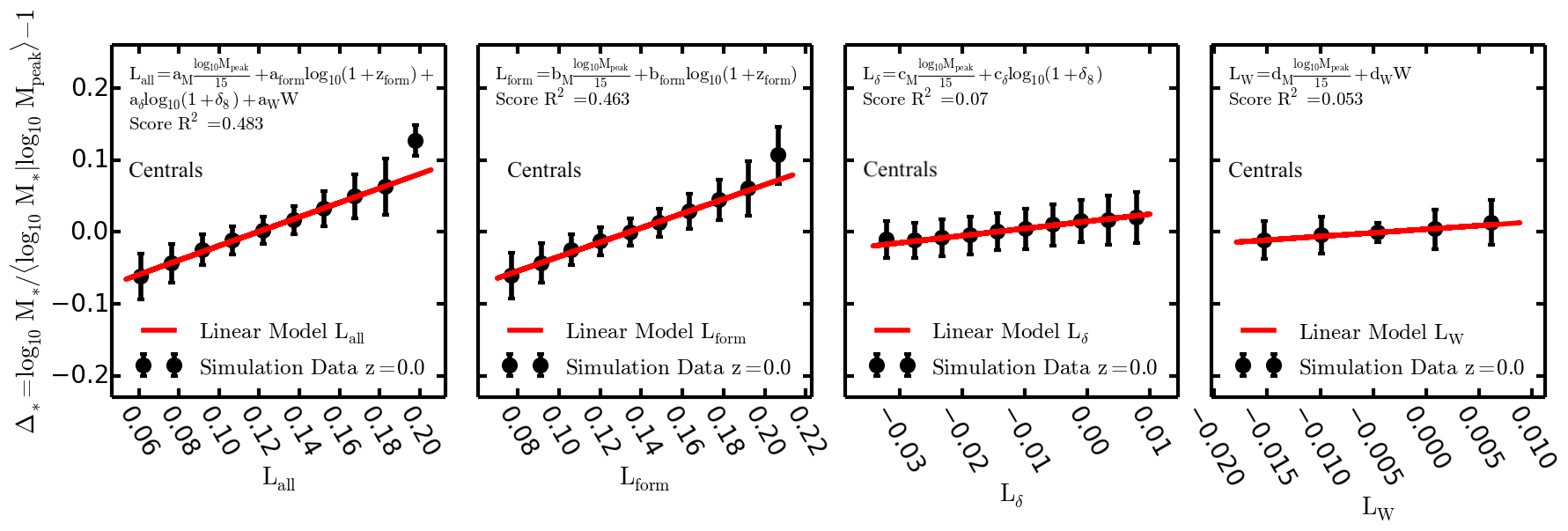}
\includegraphics[width=0.87\textwidth]{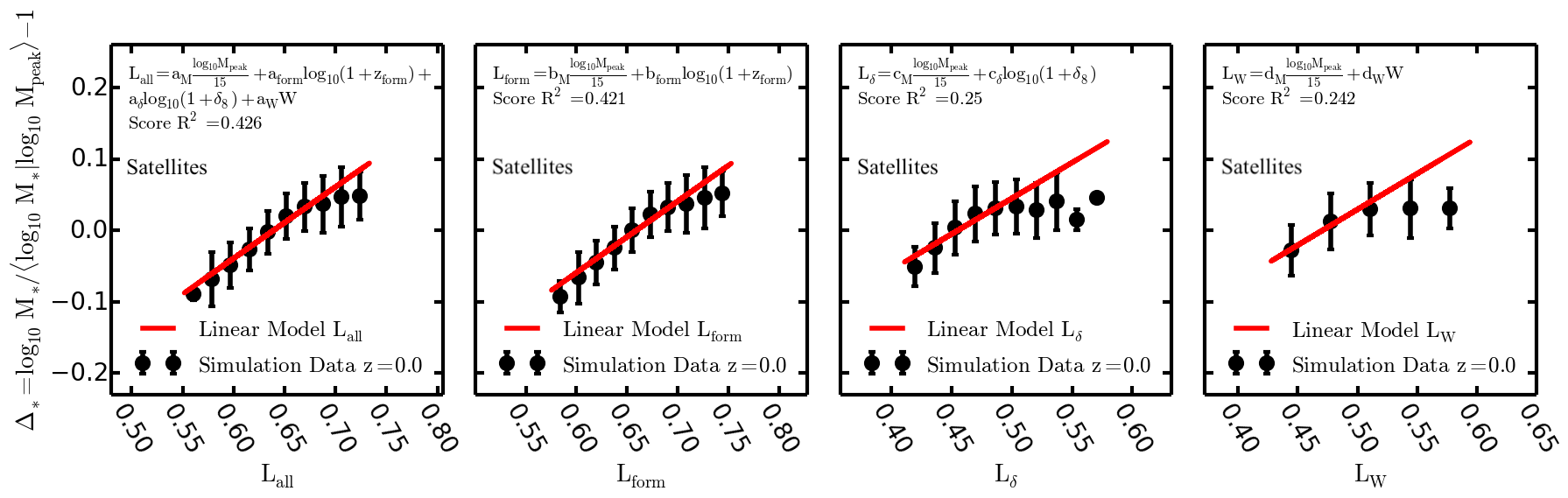}
\end{center}
\caption{Logarithmic offset $\Delta_*$ of $\log_{10}M_*$ with respect to the median measured at fixed total mass $\langle \log_{10}M_{*}|\log_{10}M_{\rm peak}\rangle$. $\Delta_*$ is plotted as as a function of the linear variables $L_{\rm all}, L_{\rm form}, L_{\rm \delta}, L_{\rm W}$ defined in equations~\ref{eq:Lall}, \ref{eq:Lform}, \ref{eq:Ldelta}, \ref{eq:LW}, respectively. The black points with error bars represent data from TNG100 (mean $\Delta_*$ and $1\sigma$ scatter within each bin). The red lines represent results of linear regressions of $\Delta_*$ with respect to the linear explanatory variables reported on the x-axes. The results are shown for central galaxies selected at redshift $z=0$ (top row) and satellite galaxies selected at redshift $z=0$ (bottom row). If only centrals are selected, the correlation of the stellar mass offset with the explanatory variables is stronger than for the whole population (top left panel). If only satellites are selected, this correlation is weaker and non-linear (bottom left panel). In particular, centrals have a stronger correlation with formation time than satellites. However, the non-linear correlation of the stellar mass offset with overdensity and Cosmic Web class is stronger for satellites than for centrals, probably as a result of environmental effects. } \label{fig:deltamstar_linear_sat_vs_cen}
\end{figure*}

Figure~\ref{fig:deltamstar_linear_sat_vs_cen} presents the result of this analysis. This figure shows that the stellar mass offset $\Delta_*$ correlates more strongly with the linear combination of all explanatory variables $L_{\rm all}$ for centrals than for the whole sample. The correlation strength of $\Delta_*$ with large-scale overdensity and Cosmic Web class, quantified by the score $R^2$, is somewhat weaker than that of the whole sample. However, the stellar mass offset $\Delta_*$ correlates more strongly with halo formation time in centrals than in the whole sample. These results indicate that centrals `carry a strong memory' of the formation site/epoch, whereas they are more weakly influenced by external effects related to the large-scale environment.

Figure~\ref{fig:deltamstar_linear_sat_vs_cen} shows that the stellar offset $\Delta_*$ of satellites behaves differently than for centrals. First of all, the relation between $\Delta_*$ and the linear combination of all explanatory variables $L_{\rm all}$ appears to be non-linear. If this relation is fitted with a linear model, the correlation score $R^2$ is a few percent smaller than for the whole sample. The non-linearity of the relation is accentuated if large-scale overdensity and Cosmic Web class are excluded from the fit. {The non-linearity appears as a strong downturn of $\Delta_*$ at high $L_{\rm all}/L_{\rm form}$, which is associated with objects with high total mass and high formation redshift ($z_{\rm form} > 0.5$)}. Nonetheless, the stellar mass offset of satellites $\Delta_*$ appears to have a stronger correlation with large-scale overdensity and Cosmic Web class than for  centrals. We interpret this result as evidence that environmental effects are able to partially `erase a galaxy's memory' of its formation site/epoch. This is not surprising, considering that many satellites may experience drastic changes of environment and environmental processes throughout their evolution. 

In summary, the stellar mass offset $\Delta_*$ is more strongly correlated with the properties and history of the underlying halo for centrals than for satellites. However, the correlation of the stellar mass offset with large-scale Cosmic Web properties is stronger for satellites than for centrals. With the use of the IllustrisTNG simulations, \cite{2018RNAAS...2....6E} and Engler et al. in prep. further quantify the deviation of the stellar to halo mass relation for satellites in groups and clusters in comparison to centrals. In agreement with, and expanding upon, the results of this work for the case of knot galaxies, they find that satellite galaxies appear to have, at fixed dynamical or total mass, enhanced stellar mass in comparison to their central counterparts.

\subsection{Low-Mass versus High-Mass Halos}\label{sec:mcut}

{In the analysis of $\Delta_*$ presented in the previous Subsections, linear regressions performed by assigning equal weights to each halo were used. Since low-mass halos are more abundant than high-mass halos, the fitting procedure is more sensitive to the former than the latter. In this Subsection, we discuss how the weighting scheme influences our results. }

\begin{figure*}
\begin{center}
\includegraphics[width=0.87\textwidth]{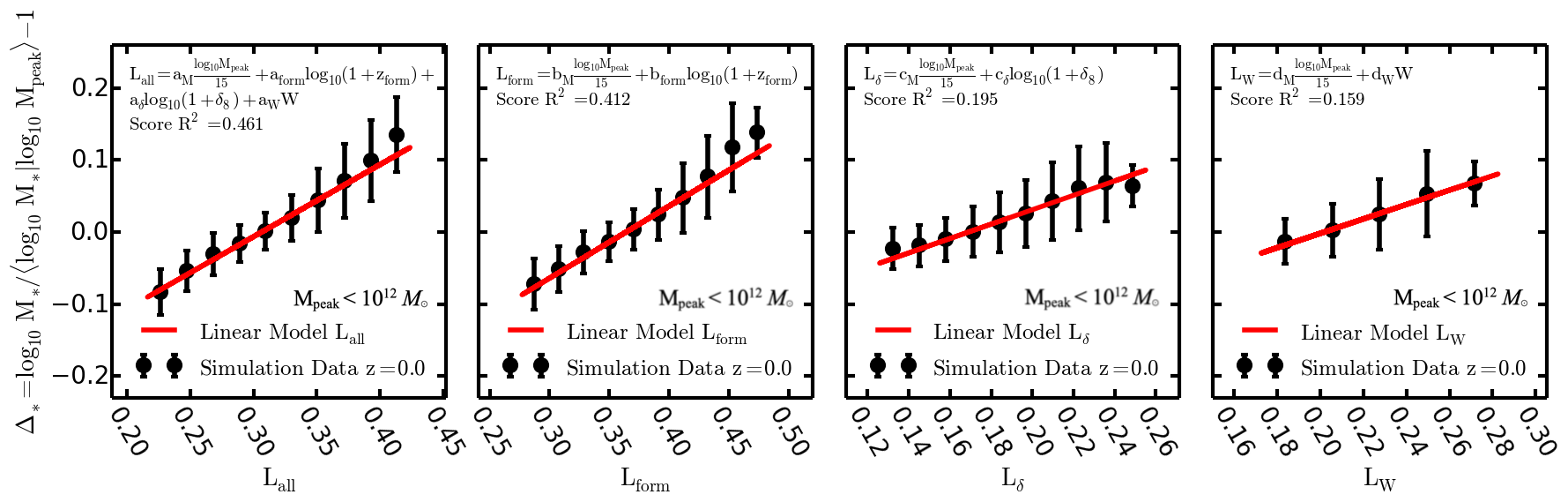}
\includegraphics[width=0.87\textwidth]{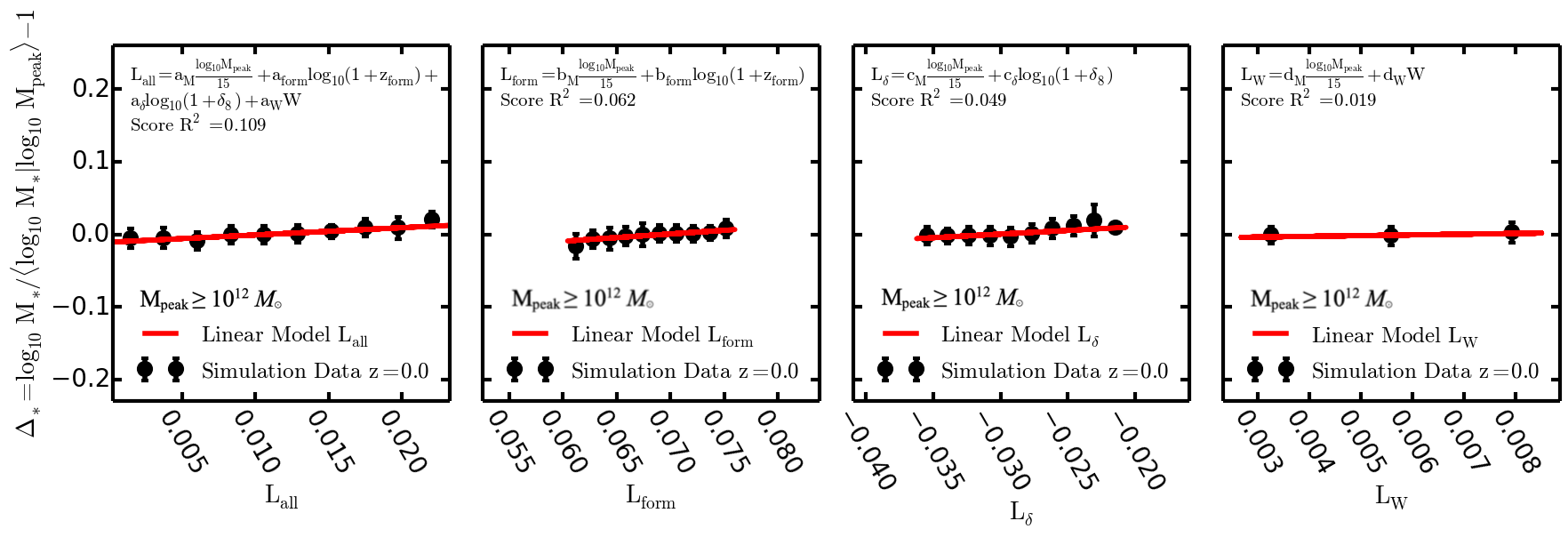}
\end{center}
\caption{{Equivalent of Figure~\ref{fig:deltamstar_linear_sat_vs_cen}, but for low-mass halos ($M_{\rm peak} < 10^{12} \, M_{\odot}$, top row) and high-mass halos ($M_{\rm peak} \geq 10^{12} \, M_{\odot}$, bottom row). The relation among $\Delta_*$ and the explanatory variables $L_{\rm all}$, $L_{\rm form}$, $L_{\rm \delta}$ and $L_{\rm W}$ is preserved for low-mass halos, but it is very weak for high-mass halos. This is a consequence of the fact that high-mass halos exhibit a small stellar mass scatter (Figure~\ref{fig:stellarmass_evo_comparison}), they typically form very early with and end up in cosmic knots, i.e. they do not exhibit a broad range of properties. Conversely, low-mass halos have a broad range of $\Delta_*$ that correlate with the formation time, large-scale overdensity and Cosmic Web class.}
} \label{fig:deltamstar_linear_low_high}
\end{figure*}

{We split the simulated galaxy sample at redshift $z=0$ in two sub-samples, one containing low-mass halos ($M_{\rm peak} < 10^{12} \, M_{\odot}$) and one containing high-mass halos ($M_{\rm peak} \geq 10^{12} \, M_{\odot}$), respectively. We repeat the linear regression procedure for each sub-sample in order to assess the robustness of our  results. Figure~\ref{fig:deltamstar_linear_low_high} shows the outcome of this test. The general conclusions of Subsection~\ref{seq:quant_res} are valid for the low-mass halo sub-sample. However, high-mass halos exhibit a very weak correlation between $\Delta_*$, total mass, formation time, large-scale overdensity and Cosmic Web class. The reason for this behavior is that very massive halos all form very early, typically end up in knots at redshift $z=0$, and have a very small scatter in stellar mass (Figure~\ref{fig:stellarmass_evo_comparison}). }

{Since the results for the low-mass halo sub-sample are the most similar to those of the total population, we conclude that low-mass halos contain the critical information to study the connection between $\Delta_*$ and the other properties considered in this paper. For this reason, we conclude that performing linear regressions with equal weighting for all galaxies does not introduce a bias that undermines the robustness of our conclusions.}

\subsection{Varying the Halo Mass Definition}\label{sec:alternative_mass_def}

{The analysis of the scatter of the relationship between halo masses and stellar masses presented above relies on a specific halo mass definition, the peak mass reached by any given halo throughout its evolution. Although this quantity is of great theoretical relevance, it is not measurable. For this reason, the analysis was repeated by adopting an alternative definition of the halo mass that is similar to total masses determined from observations: $M_{\rm sub}$, the mass of a gravitationally bound object as identified by {\sc subfind} and including a priori all matter components: dark matter, gas and stars. The `sub' subscript refers to the fact that we use masses computed by {\sc subfind}. This can be intended as the dynamical mass of a self-gravitating (sub)halo or galaxy at a given time. }

\begin{figure*}
\begin{center}
\includegraphics[width=0.87\textwidth]{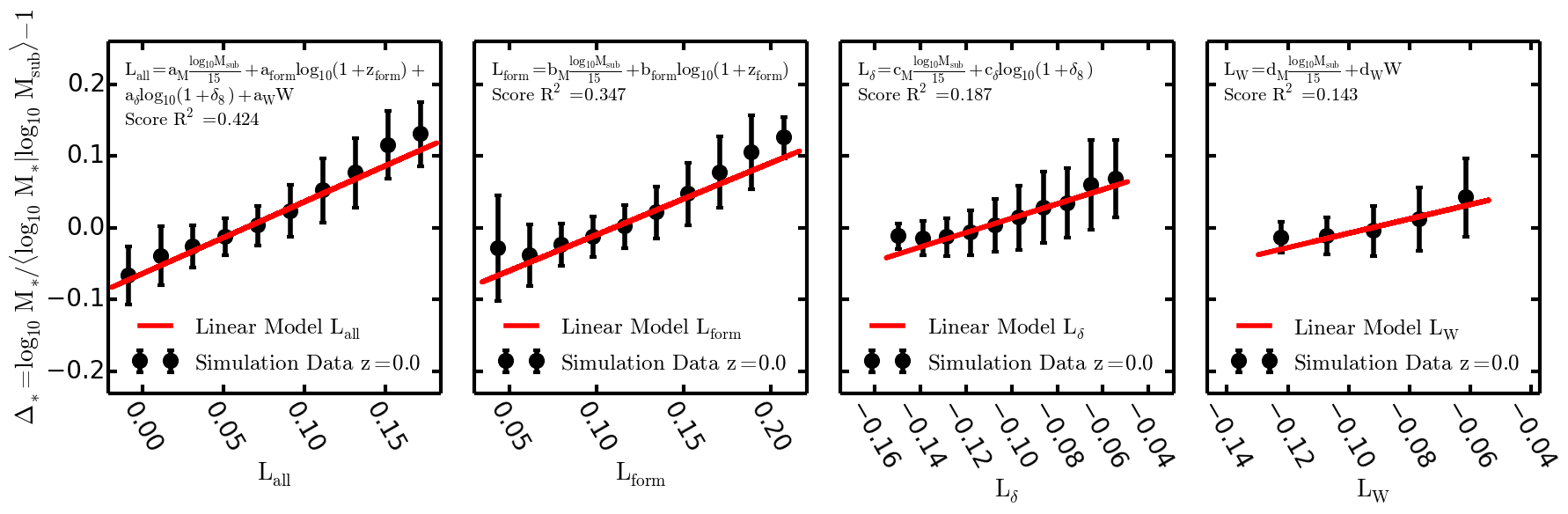}
\end{center}
\caption{{Equivalent of Figure~\ref{fig:deltamstar_linear}, but for a different definition of the halo mass: $M_{\rm sub}$, the mass of a gravitationally bound object as identified by {\sc subfind}, which is similar total masses determined from observations. The principal results of the analysis of the scatter of the relationship between halo masses and stellar masses are confirmed with this alternative halo mass definition.}
} \label{fig:deltamstar_linear_alt}
\end{figure*}

{Figure~\ref{fig:deltamstar_linear_alt} shows the result of the linear regressions of $\Delta_* = \log_{10}M_{*}/\langle \log_{10}M_{*}|\log_{10}M_{\rm sub}\rangle-1$ against the explanatory variables $L_{\rm all}$, $L_{\rm form}$, $L_{\rm \delta}$ and $L_{\rm W}$ (equations~\ref{eq:Lall}-\ref{eq:LW}), where $M_{\rm peak}$ was replaced with $M_{\rm sub}$. Choosing this alternative mass definition worsens the quality of the linear fits, and $R^2$ is a few \% smaller than in the fiducial case. This difference is likely caused by the fact that the value of $M_{\rm sub}$ is influenced by the non-linear and stochastic history of environmental and stripping processes experienced by satellite halos, which is not the case when using $M_{\rm peak}$. Apart from this minor difference, the results are in qualitative agreement with those found in the fiducial analysis, and the interpretation is the same as the one discussed in   Subsections~\ref{seq:quant_res}-\ref{sec:mcut}. This conclusion is also confirmed by noting the similarity of the coefficients of the alternative linear regression models summarized in Table~\ref{tab:params_alt} with the ones of the fiducial models in Table~\ref{tab:params}.}

\begin{table}
\centering
\caption{ {Linear regression coefficients for the scaling of the logarithmic offset of galaxy stellar masses $\Delta_*$ as a function of (normalised) total mass $(\log_{10}M_{\rm sub})/15$, halo formation redshift $\log_{10}(1+z_{\rm form})$, large-scale overdensity $\log_{10}(1+\delta_{8})$ and Cosmic Web class $W$ at multiple redshifts. See equations~\ref{eq:Lall}-\ref{eq:LW}.}}\label{tab:params_alt}
{\bfseries Alternative linear model coefficients for $\Delta_*$}
\makebox[\linewidth]{
\begin{tabular}{lrrrr}
\hline
\hline
\multicolumn{5}{l}{Model $\Delta_*(L_{\rm all})$} \\ 
\hline
 Parameter & $z=0$ & $z=1$ & $z=2$ & $z=4$ \\
\hline
 $\Delta_{\rm all}$ & -0.0176 & -0.1965 & -0.3264 & -0.5841 \\ 
 $a_{\rm M}$ & -0.0835 &  0.0372 &  0.0578 &  0.1152 \\ 
 $a_{\rm form}$ &  0.1798 &  0.3160 &  0.4539 &  0.6427 \\ 
 $a_{\rm \delta}$ &  0.0282 &  0.0077 &  0.0025 & -0.0003 \\ 
 $a_{\rm W}$ &  0.0095 &  0.0203 &  0.0139 &  0.0023 \\ 
 \hline
 \hline
\multicolumn{5}{l}{Model $\Delta_*(L_{\rm form})$} \\ 
\hline
 Parameter & $z=0$ & $z=1$ & $z=2$ & $z=4$ \\
\hline
 $\Delta_{\rm form}$ & -0.1098 & -0.2328 & -0.3469 & -0.5874 \\ 
 $b_{\rm M}$ &  0.0472 &  0.0931 &  0.0917 &  0.1205 \\ 
 $b_{\rm form}$ &  0.2293 &  0.3335 &  0.4611 &  0.6432 \\
 \hline
 \hline
\multicolumn{5}{l}{Model $\Delta_*(L_{\rm \delta})$} \\ 
\hline
 Parameter & $z=0$ & $z=1$ & $z=2$ & $z=4$ \\
\hline
 $\Delta_{\rm \delta}$ &  0.1539 &  0.0928 &  0.0733 &  0.0109 \\ 
 $c_{\rm M}$ & -0.2335 & -0.1352 & -0.1077 & -0.0121 \\ 
 $c_{\rm \delta}$ &  0.0440 &  0.0217 &  0.0120 & -0.0016 \\
 \hline
 \hline
\multicolumn{5}{l}{Model $\Delta_*(L_{\rm W})$} \\ 
\hline
 Parameter & $z=0$ & $z=1$ & $z=2$ & $z=4$ \\
\hline
 $\Delta_{\rm W}$ &  0.1180 &  0.0837 &  0.0752 &  0.0167 \\ 
 $d_{\rm M}$ & -0.2080 & -0.1388 & -0.1184 & -0.0254 \\ 
 $d_{\rm W}$ &  0.0760 &  0.0445 &  0.0274 & 0.0044 \\ 
\hline
\hline
\end{tabular}
}
\end{table}

\section{Discussion and Conclusions} \label{sec:conclusions}

In this paper, we have studied the connection among stellar mass, total mass, halo formation time and  the location in the large-scale Cosmic Web of galaxies using the data from the TNG100 large-volume cosmological hydrodynamical simulation. The analysis is based on the Cosmic Web classification method we already used in \cite{2019MNRAS.486.3766M}, which uses the deformation tensor of the total matter density field smoothed with a Gaussian kernel of radius $R_{\rm G} = 8 \, h^{-1}{\rm Mpc}$. This classification separates the large-scale cosmological Cosmic Web from structure at smaller scales, such as halos, which are identified with the {\sc subfind} algorithm. We select objects with total dynamical mass $\geq 6.3\times 10^{10} h^{-1}\, M_{\odot}$ and up to a few $10^{14} h^{-1} \, M_{\odot}$ between redshift $z=4$ and redshift $z=0$. For each redshift and for each object in the IllustrisTNG group catalogues, we measure its total mass, formation redshift, value of the local large-scale overdensity and Cosmic Web class (knot, filament, sheet, void). We show that the local large-scale overdensity and the Cosmic Web class are not entirely degenerate variables, because the latter includes information about the density field morphology which is not contained in the former. 

We confirm that galaxy stellar mass strongly correlates with total mass and formation time, and we uncover that it correlates more weakly with large-scale overdensity and Cosmic Web class. 
In particular, we find that the scatter in the total mass versus stellar mass relation correlates with both the large-scale overdensity and with the Cosmic Web class. Our quantitative analysis of simulation data shows that this is not a small effect, and that up to $\sim 20\%$ of this scatter at redshift $z=0$ can be explained with a deterministic linear combination of total mass, overdensity and Cosmic Web class as variables. If formation redshift is included as a variable of the model, the latter can explain up to $\sim 40\%$ of the scatter. Similar conclusions apply to higher redshift, but the role of overdensity and Cosmic Web class becomes weaker (only $\sim 5\%$ of the scatter). 

These results suggest that {\it on average} the earlier an overdensity develops, the earlier halos/galaxies will form within it, and the more time these halos/galaxies will have to fall into a knot or filament, and to assemble a higher-than-average stellar mass. In lower density regions such as sheets and voids, halos/galaxies {\it on average} form later, and have less time to accumulate dark matter and form stars. The correlations found in this work are mostly determined by central galaxies. Indeed, the fact that the correlation between stellar mass, halo mass, formation time and large-scale environment is not perfect, implies that this simplistic scenario may have exceptions. For instance, satellites may form late in a sheet or void, then migrate to a knot and lose part of their mass due to stripping processes (see Engler et al. in prep.). The fact that these exceptions exist, is supported by our finding that Cosmic Web class/large-scale overdensity have a weak correlation with formation time (see Table~\ref{tab:covariance}).

Driven by these considerations, we further extended our analysis to differentiate between central and satellite galaxies. In qualitative terms, the correlations discussed above also exist for these two classes. However, the stellar masses of centrals are found to correlate more strongly with the properties of their halos than satellites, whereas satellites appear to have a higher relative correlation strength with large-scale overdensity and Cosmic Web class. We interpret this result as a consequence of satellites being perturbed by environmental effects that are triggered when they transit from one location of the Cosmic Web to another. A typical example is a galaxy streaming along a filament until it falls into a knot, and is then accreted by a large galaxy cluster, where it experiences stripping processes. Galaxies that manage to survive as centrals of a halo are shielded by these environmental effects and manage to retain the dependence on the conditions at the location/epoch where/when they formed. 

The procedure we followed to study the connection between stellar masses, total dynamical masses and the large-scale cosmic density field is robust, but difficult to apply directly to galaxy survey for a series of reasons. Although motivated by robust theoretical arguments, the Cosmic Web classification method we use is based on knowledge of the total large-scale density field, and cannot be straightforwardly applied to survey data. Nonetheless, a similar implementation, the {\sc DisPerSE} code of \cite{2011MNRAS.414..350S}, has been used by \cite{2019MNRAS.483..172D} to connect MaNGA galaxy kinematics to assembly history. An implementation of the deformation tensor method we used was also recently used on a sample of $\sim 10^5$ galaxies from the GAMA survey \citep{2015MNRAS.448.3665E,2016MNRAS.462.4451B}. This analysis lead to the conclusion that the galaxy mass function does not depend on the deformation tensor, but only on the large-scale overdensity. However, the robustness of these results is somewhat limited by the size of the sample, and by the fact that the classification method is directly applied to the galaxy overdensity field which is biased with respect to the total density field. An additional limitation to fully measure the effects discussed in this paper, is the fact that the halo mass of each galaxy should be also known. In practice, simultaneously obtaining the halo mass (e.g. from gravitational lensing) and information on the large-scale density field from galaxy surveys is challenging. 

Nonetheless, multiple cutting-edge methods to measure the Cosmic Web structure directly from surveys have been designed \citep{2011MNRAS.414..350S,2018MNRAS.473.1195L}, and, in principle, it should be possible to apply them and verify whether some of the effects we identify are seen in the real Universe. For instance, one could perform a Cosmic Web classification on a large galaxy redshift survey, select galaxies in a given stellar mass range, and then compute the median stellar mass for each Cosmic Web environment. If our results hold, the median stellar mass should depend on the Cosmic Web class.

In conclusion, our analysis suggests the existence of non-trivial connection between galaxies, their halos and the large-scale environment that can be measured in the real Universe, with currently available data, and in the near future. On the theoretical side, there are a number of ways in which the present analysis can be improved: the first one is the use of Bayesian methods to study the correlations we identified in this work \citep[e.g.][]{2008ConPh..49...71T}, which would allow us to penalize certain analytical models against others; the second one is a detailed study of what causes the correlations, which would involve tracking the flow of dark matter and baryons on a halo-to-halo basis; the third one is the employment of Cosmic Web classification tools that can be directly applied to galaxy survey data, and that are weakly affected by galaxy biasing. We defer these improvements to future work.

\section*{Acknowledgments}
DM was supported by the DARK Fellowship. DM and SH acknowledge contribution from the Danish council for independent research under the project DFF-6108-00470, and from the Danish National Research Foundation project DNRF132. PT acknowledges support from NASA through Hubble Fellowship grant HST-HF2-51341.001-A awarded by STScI, which is operated under contract NAS5-26555. FM acknowledges support through the program ``Rita Levi Montalcini'' of the Italian MIUR. The IllustrisTNG simulations and the ancillary runs were run on the HazelHen Cray XC40-system (project GCS-ILLU), Stampede supercomputer at TACC/XSEDE (allocation AST140063), at the Hydra, Draco supercomputers at the Max Planck Computing and Data Facility, and on the MIT/Harvard computing facilities supported by FAS and MIT MKI. We also thank the Reviewer and Editors of this paper for their useful comments.
\bibliography{main}

\end{document}